\documentstyle[12pt,epsfig]{article}
\newcommand{\be}{\begin{equation}}
\newcommand{\bea}{\begin{eqnarray}}
\newcommand{\eea}{\end{eqnarray}}
\newcommand{\ba}{\begin{array}}
\newcommand{\ea}{\end{array}}
\newcommand{\ee}{\end{equation}}

\newcommand{\cN}{{\cal N}}

\expandafter\ifx\csname mathbbm\endcsname\relax

\else

\fi
\def\a{\alpha}

\begin{document}
\begin{titlepage}
\hfill
\vbox{
    \halign{#\hfil         \cr
           CERN-TH/2000-222 \cr
           hep-th/0007215   \cr
           } 
      }  
\vspace*{20mm}
\begin{center}
{\Large {\bf  Supergravity and Light-Like Non-commutativity}\\} 

\vspace*{15mm}
\vspace*{1mm}
{Mohsen Alishahiha$^1$, Yaron Oz$^1$ and Jorge G. Russo$^2$}\\

\vspace*{1cm} 

{\it $^1$ Theory Division, CERN \\
CH-1211, Geneva, 23, Switzerland}\\
\bigskip
{\it $^2$ Departamento de F\'\i sica, Universidad de Buenos Aires\\
Ciudad Universitaria, 1428 Buenos Aires}\\

\vspace*{1cm}
\end{center}

\begin{abstract}

We construct dual supergravity descriptions of field theories 
and little string theories with 
light-like non-commutativity. The field theories
are realized on the world-volume of Dp branes with light-like NS $B$ field
and M5 branes with light-like
$C$ field. The little string theories are  realized on the world-volume of
NS5 branes with light-like RR $A$ fields. The supergravity backgrounds are closely
related to the $A=0,B=0,C=0$ backgrounds. We discuss the implications
of these results.
We also construct dual supergravity descriptions of ODp theories realized on
the worldvolume of NS5 branes with RR backgrounds.

\phantom{\cite{CDS,HulDo,sw,sst1,GMMS,br,gm,GMSS,MR,AOS}}

\phantom{\cite{sst,gars,BIO,H,km,bb,chen,rj,kawt,bs,rv,har,ohta,LRS}}

\end{abstract}
\vskip 5cm

July 2000
\end{titlepage}

\newpage

\section{ Introduction}
Non-commutative field theories, open string theories and open 
membrane theories
have been extensively studied recently.
These theories are realized on the world-volume of Dp branes with a nonzero 
NS $B$ field and  M5 branes with nonzero $C$ field.

While space non-commutativity can be accommodated within field theory
\cite{CDS,HulDo,sw}, 
space-time non-commutativity seems to require string theory for consistency
\cite{sst1}--\cite{GMSS}.\footnote{
Dual supergravity descriptions of these theories
have been  constructed in 
\cite{MR,AOS}. Further interesting progress in the study of
theories with space-time non-commutativity
has been made in \cite{sst}--\cite{LRS}.}
However, it was argued in \cite{agm} that light-like non-commutativity
(that is, $[x^\mu,x^\nu ]=i\theta^{\mu\nu}$, where $\theta^{\mu\nu}$
is light-like, e.g. $\theta^{+2}\neq 0$) can be realized within field theories.
One of the aims of this paper is to
construct dual supergravity descriptions of such field theories 
in various dimensions.
The field theories
are realized on the world-volume of Dp branes with light-like NS $B$ field
in a particular decoupling limit.

Of particular interest is a six dimensional field theory with light-like
 noncommutativity
realized on the world-volume  of M5 branes with light-like
$C$ field. This theory is conjectured to have a DLCQ matrix description
as a quantum mechanics on the resolved moduli space of instantons 
with  the  light-like
$C$ field corresponding to the resolution parameter.
We will construct a dual supergravity description of this theory.
We will see that the supergravity descriptions of 
field theories  with light-like non-commutativity are closely related
to those without non-commutativity, and we will discuss the implications.

Another class of interesting theories are the non-commutative little
string theories.
These theories are  realized on the world-volume of
NS5 branes with light-like RR $A$ fields.
We will 
construct dual supergravity descriptions of these theories. 

A second aim of this paper is to construct  dual supergravity descriptions
of theories realized on the  worldvolume  of NS5 branes, whose 
excitations include light-open Dp branes (ODp) \cite{{GMSS},{har}}.
Such backgrounds are obtained from NS5 branes with near critical
RR fields. 

The paper is organized as follows.
In section 2 we  
construct Dp brane,   M5 brane and NS5 brane solutions  
corresponding to theories 
in various dimensions with 
light-like non-commutativity. 
We also discuss some salient aspects of these solutions 
and  what we learn from them about the corresponding  field theories.
In section 3 we  construct  supergravity  solutions
of  NS5 branes with RR backgrounds and obtain
dual supergravity descriptions of ODp theories. 
As an application, we compute the absorption cross section of a graviton
polarized along the world-volume.

\section{Branes with light-like background  fields}

\subsection{Construction of D-brane solutions}
\def\a{\alpha }
\def\g{\gamma }
\def\del{\partial }
\def\s{\sigma}

Our aim is  to construct  supergravity 
backgrounds with Dp brane charge and a $B$-field
with components $B_{-2}$.
We will start with the D3 brane case.
The construction can be done by performing
an infinite Lorentz boost in the $x^1$ direction on
the known \cite{MR,hashi}  D3 brane
background in the presence of a $B_{12}$ field.
A finite Lorentz boost gives the following background
\be
ds^2=f^{-{1\over 2}}\big[-d\tilde x_0^2 +dx_3^2+ {f\over H}(d\tilde x_1^2+dx_2^2)\big]
+f^{{1\over 2}}\big( dr^2 +r^2 d\Omega_5^2\big)\ ,
\label{mttt}
\ee
$$
f=1+{ {\a'}^2 R^4\over r^4}\ ,\ \ \ \ \ H=1+{{\a'}^2 R^4\over r^4}\cos^2\a\ ,
$$
$$
e^{2\phi}=g_s^2 {f\over H} \ ,\ \  \ \ 
F_{\tilde 0 \tilde 1 2 3r}={1\over g_s}\cos\a {f\over H}\del_r f^{-1}\ ,
$$
\be
B=\tan\a \ H^{-1} \ d\tilde x_1\wedge dx_2\ ,\ \ \ \ \ \ 
A={1\over g_s} \sin\a \ f^{-1} \ d\tilde x_0\wedge dx_3\ ,
\label{aaa}
\ee
where
\be
\tilde x_0=\cosh\g \ x_0-\sinh\g \ x_1\ ,\ \ \ \ \ 
\tilde x_1 =-\sinh\g \ x_0 +\cosh\g \ x_1\ ,
\label{xxx}
\ee
or $\tilde x_+=e^{-\g }x_+\ ,\ \ \tilde x_-=e^{\g} x_-$, with
$x_\pm =\pm x_0+x_1\ $, and $A$ is the RR 2-form.
Note also that $F_{\tilde 0 \tilde 1 2 3r}=F_{ 0 1 2 3r}$.

We obtain
$$
ds^2=f^{-{1\over 2}}\big[-dx_0^2 +dx_1^2+ dx_3^2+ {f\over H}
d x_2^2 +   {{\a'}^2 R^4\over Hr^4}\sin^2\a  
(\cosh\g \ dx_1-\sinh\g \ dx_0 )^2 \big]
$$
\be
+f^{{1\over 2}}\big( dr^2 +r^2 d\Omega_5^2\big)\ ,
\ee
$$
e^{2\phi}=g_s^2 {f\over H} \ ,\ \  \ \ 
F_{0 1 2 3r}={1\over g_s}\cos\a {f\over H}\del_r f^{-1}\ ,
$$
\be
B_{02}=-\tan\a \sinh\g \ H^{-1} \ ,\ \ \ \ \ \ 
B_{12}=\tan\a \cosh\g \ H^{-1} \ ,
\label{aab}
\ee
\be
A_{03}={1\over g_s} \sin\a \cosh\g \ f^{-1} \ ,
\ \ \ 
A_{13}= - {1\over g_s} \sin\a \sinh\g \ f^{-1}\ .
\label{aabb}
\ee

The fields take the following asymptotic values
\be
B^\infty_{02}=-\tan\a\sinh\g \equiv E\ ,\ \ \ \ 
B^\infty_{12}= \tan\a\cosh\g \equiv B\ ,\ \
\label{asy}
\ee
$$
A^\infty_{03}={1\over g_s} \sin\a\cosh\g ={\sqrt{B^2-E^2}\cosh\g\over
g_s\sqrt{1+B^2-E^2}}\ ,\ \ \ 
$$
$$
A^\infty_{13}=-{1\over g_s} \sin\a\sinh\g =-{\sqrt{B^2-E^2}\sinh\g\over
g_s\sqrt{1+B^2-E^2}}\ ,
$$
where we have used $B^2-E^2=\tan^2\a $.
To have only light-like B-field components, we now take the infinite boost limit,
$\gamma \to\infty $. At the same time, we must take the limit $\a\to 0 $ with
$$
e^\g\tan\a={\rm finite }\equiv b \ .
$$
In this limit the asymptotic values of the gauge fields simply become
\be
B_{02}^\infty =E=-b\ ,\ \ \ B_{12}^\infty =B=b\ ,\ \ \ \ A_{03}^\infty ={b\over g_s}=-A_{13}^\infty \ .
\label{ggg}
\ee
We obtain the background  
\be
ds^2=f^{-{1\over 2}}\big[dx_+dx_- + d x_2^2+dx_3^2
+{ {\a'}^2 R^4b^2\over r^4f} dx_-^2\big]
+f^{{1\over 2}}\big( dr^2 +r^2 d\Omega_5^2\big)\ ,
\label{mtt}
\ee
$$
e^{2\phi}=g_s^2  \ ,\ \  \ \ 
F_{ 0 1 2 3r}={1\over g_s}\del_r f^{-1}\ ,
$$
\be
B= b f^{-1} \ d x_-\wedge dx_2\ ,\ \ \ \ \ \ 
A=-{1\over g_s} b f^{-1} \ d x_-\wedge dx_3\ .
\label{zzz}
\ee
Note that this is a constant dilaton solution representing a wave travelling
on the D3 brane, but with $B$ and $A$ fields.
The same background is obtained by a similar procedure
by starting with the purely electric field configuration $B_{03}\neq 0$ and performing a Lorentz boost. An alternative derivation by dualities
is described in   subsection 2.3.

Let us  consider the decoupling limit $\a'\to 0 $. 
Before taking the limit, it is convenient to  
redefine the coordinate $x_+\to x_+ - { b^2} x_- $.
The metric (\ref{mtt}) takes the form
\be
ds^2=f^{-{1\over 2}}\big[dx_+dx_- + d x_2^2+dx_3^2
-{ b^2\over f} dx_-^2\big]
+f^{{1\over 2}}\big( dr^2 +r^2 d\Omega_5^2\big)\ ,
\label{mmtt}
\ee
We set as usual $r=\a' u$, with $u$ fixed.
In addition, in order to have non-vanishing gauge fields $B$ and $A$
after $\a'\to 0$, we must rescale $b$ by introducing
$\tilde b=\a' b$= fixed.
So $u, R, \tilde b,g_s$ remain fixed.
We get the following background
$$
ds^2= \a' \bigg({u^2\over R^2} \big[ dx_-dx_+ + d x_2^2+dx_3^2 
-{\tilde b^2\over R^4} u^4 dx_-^2\big]
+ R^2{du^2 \over u^2}+  R^2d\Omega_5^2 \bigg)\ ,
$$
\be
e^{2\phi}=g_s^2  \ ,\ \  \ \ 
F_{ 0 1 2 3r}={\a'}^2{4\over g_s R^4} u^3 \ , \ \ \ R^4=4\pi g_s N=2g_{\rm YM}^2 N\ ,
\label{deco}
\ee
$$
B= \a' \tilde b {u^4\over R^4} \ d x_-\wedge dx_2\ ,\ \ \ \  
A=-\a' {\tilde b\over g_s}  {u^4\over R^4} \ d x_-\wedge dx_3\ ,
\ \ \ \ x_\pm=x_1 \pm x_0 \ .
$$
Note that $g_s$ was maintained fixed and $ b\to\infty $ in the decoupling limit,  in accordance with the field theory analysis of \cite{agm}.

The above solutions can be easily generalized to the case of the Dp branes,
with $p=2,...,5$.
A similar procedure gives  the following solution representing
a wave on the Dp brane:
\bea
ds^2&=&f^{-{1\over 2}}\big[dx_+dx_- + d x_2^2+...+dx_p^2
-{ b^2\over f} dx_-^2\big]
+f^{{1\over 2}}\big( dr^2 +r^2 d\Omega_{8-p}^2\big)\ ,\cr
e^{2\phi}&=&g_s^2 f^{{3-p\over 2} },\;\;\;\;\; 
 f=1+  {c_p{\a'}^{5-p}g^2_{\rm YM} N \over r ^{7-p}},\;\;\;\;\;
c_p=2^{7-2p} \pi^{ {9-3p\over 2} } \Gamma
\big( {7-p\over 2} \big),\cr
B_{-2}&=& b f^{-1},\;\;\;\;\;\;\;A^{(p-1)}_{-3\cdots}=-{b\over g_s}f^{-1}
\label{bbk}
\eea
where $A^{(p-1)}$ is an RR $(p-1)$-form. Note that for D2-brane we have
a one-form $A^1_{-}$. There is also the usual RR form which gives  the
 Dp-brane charge. The gauge coupling is
\be
g^2_{\rm YM}=(2\pi )^{p-2} g_s\ {\a'}^{ {p-3\over 2} } \ . 
\ee

The decoupling limit is obtained by rescaling variables as follows:
$$
r=\a' u\ , \ \  \ \ 
\tilde b=\a' b \ , \ \ \ \
$$
and taking the limit $\a'\to 0$ with  $u, \tilde b, \ g^2_{\rm YM}$ fixed.

We get 
$$
ds^2= \a'   {u^{{p-3\over 2}} \over \sqrt{\lambda } }
 \bigg(  {u^{5-p}} \big[ 
dx_-dx_+ + d x_2^2+...+dx_p^2 
-\tilde b^2 { u^{7-p}\over \lambda } dx_-^2 \big]
+  \lambda  {du^2 \over u^2}+  \lambda d\Omega_{8-p}^2 \bigg)\ ,
$$
$$
e^{\phi}= g_{\rm YM}^2 \ { 
 u^{ {1\over 4} (p-3)(7-p) }  
\over (2\pi )^{p-2}  
\lambda ^{ {1\over 4}(p-3)} }    =
{1\over N}\ {(\lambda u^{p-3})^{ {1\over 4}(7-p)}\over c_p(2\pi)^{p-2} }
  \ ,\ \  \ \ \lambda \equiv c_p g_{\rm YM}^2 N \ ,
$$
\be
B= \a' \tilde b \ {u^{7-p}\over \lambda } \ d x_-\wedge dx_2\ .\ \ \ \  
\label{bba}
\ee

\subsection{ Light-Like Non-commutative SYM}

The Dp brane background  (\ref{bba})
should provide a dual supergravity description of
the $p+1$ dimensional light-like
non-commutative super Yang-Mills theory, where
$\theta^{\mu\nu}$ has only  non-vanishing $\theta^{+2}=-\theta^{2+}$
components.
Since the parameter $\tilde b$ can be scaled away from the metric by
$x_-\to x_-/\tilde b$, $x_+\to x_+\tilde b $, the curvature invariants are independent of $\tilde b$. They are functions only of the coordinate $u$,
and therefore they must be the same as in the (commutative) $\tilde b=0$
case. Thus the regime of validity of supergravity approximation
is as in the commutative case.

The curvature of the metric in eq.~(\ref{bba}) has the behavior 
\be 
l_s^2{\cal R}\sim \frac{1}{g_{\rm eff}}\ ,
\ee
where we have defined a dimensionless effective gauge coupling $g_{\rm eff}$ 
\be
g_{\rm eff}^2 = g_{\rm YM}^2\; N u^{p-3} \ .
\ee
As usual, when $g_{\rm eff}\ll 1$ the perturbative field theory description is 
valid, while when $g_{\rm eff} \gg 1$ the dual supergravity description is valid. 
The expressions for the dilaton and curvature thus indicate that the phase structure
of the non-commutative light-like theory is similar to that of the ordinary 
Dp-branes.

Let us now consider the case $p=3$ in detail.
The D3 brane background (\ref{deco}) provides a dual description of
$3+1$ dimensional  super Yang-Mills theory
with light-like non-commutativity.
The solution (\ref{deco})  is the first example of a constant
dilaton solution describing a non-commutative field theory.
Remarkably, this geometry   has
also constant curvature invariants,
e.g.  ${\cal R}$, ${\cal R}_{\mu\nu}{\cal R}^{\mu\nu}$,
and ${\cal R}_{\mu\nu\rho\s }^2$, which have just the same values as
in the $AdS_5\times S^5$ case. The fact that they are constants can be understood
from the invariance  of the metric  
under the scaling $u\to c\ u$, with an appropriate rescaling of $x_+,x_-,x_2,x_3$. 
In particular,  although the Weyl tensor has some non-vanishing components, 
the square Weyl tensor $C_{\mu\nu\rho\lambda }^2$ is identically zero.
Along with the fact that the dilaton is constant, this suggests a mild energy dependence of the gauge coupling.
In the low energy region $u\cong 0$, the metric approaches $AdS_5\times S^5$, which is consistent with the expectation that the low-energy
theory should be described by the usual ${\cal N}=4$ SYM theory.

For $g_s\gg 1$, the dilaton is large and we have to use the S-dual picture. 
The S-dual background is simply obtained by $g_s\to 1/g_s$ and exchanging
the gauge fields $B$ and $A$, which gives  $B_{-3}$ and $A_{-2}$ non-zero components.
Therefore the 
strong coupling limit of 
SYM theory with light-like non-commutativity 
is another
NCSYM theory with 
non-commutativity in light-like directions, but with $\theta^{+3}$ instead of $\theta^{+2}$. The exchange of 2-3 directions can be undone
by an $SO(2)$ rotation in the plane $x_2$-$x_3$, 
since  light-like  NCSYM theory is invariant under 
such $SO(2)$ transformations (see also \cite{LRS}).
More general $SL(2, R)$ transformations  only mix the
$B_{-2}$ and $A_{-3}$ components and  introduce a constant RR scalar
field $\chi $ (which  in the 
low-energy Yang-Mills theory gives rise to  a $\Theta $ term $\Theta\int \tilde F F$).

The supergravity background  (\ref{deco})
is notably simple,
in fact, given by a simple perturbation of the $AdS_5\times S^5$
background:
\be
g_{\mu\nu}\to g_{\mu\nu}+\delta g_{\mu\nu}\ ,\ \ \ 
\delta g_{--}=-\tilde b^2\ {u^6\over R^6} \ ,
\label{yyy}
\ee
\be
\delta B_{-2}=\tilde b \ {u^4\over R^4}\ ,\ \ \ \ 
\delta A_{-3}= - {\tilde b\over g_s} \ {u^4\over R^4}\ .\ \ 
\label{zzx}
\ee
The form of the perturbations (\ref{yyy}), (\ref{zzx}) implies
that as an expansion
in the non-commutativity parameter $\tilde b$ the first-order perturbation
around the $\cN=4$ background is exact. Note in comparison that
for  spatial non-commutativity, say in coordinates $x_2,x_3$, there are
infinite number of terms in the   expansion
in the non-commutativity parameter around the $\cN=4$ background.
This 
suggests that there could exist a simple
modification of the  ${\cal N}=4$ SYM theory Lagrangian
which turns it into a 
 light-like non-commutative SYM theory.
The pure gauge theory part of
the dimension six SCFT operator corresponding to $\delta B_{-2}$ is 
\cite{das}
\be
{\cal O}_6=
\left[ F^{2m}F_{mk} F^{k-}
+{1\over 4} F_{lm}F^{ml} F^{-2}\right] \ .
\label{oop}
\ee
A question of interest is whether  the resulting theory, after adding 
a perturbation of the form ${\cal O}_{6}\delta B_{-2}$ to the SYM action, 
is exactly equivalent to the  non-commutative field theory action.
This is not the case and one can check that with 
 light-like non-commutativity there are still infinite number of terms.
In particular, in the abelian case one can compute the NCSYM action using Seiberg-Witten map \cite{sw}, and show that the Lagrangian contains
arbitrary powers in $\theta^{+2}$.\footnote{However,
it is interesting to note that in the particular case
that $F_{-2}=0$, 
the series truncates and the full Lagrangian
contains only the usual SYM action plus a linear term in $\theta^{+2}$.}

Thus, while in the perturbative SYM description
there are an infinite number of terms, there seems to be a  simplification 
at strong  t' Hooft coupling $g_sN$.
The S-duality symmetry of the theory is not useful in this limit,
since the simplification seems to take place at strong 't Hooft coupling
$g_sN$ and large $N$, but $g_s\ll 1$.
A possible explanation for the simplification at strong coupling
is an existence of a simple resummation
of the perturbation series. 

The strong coupling expansion is different from that of the ordinary 
$\cN=4$ SYM theory, 
since the geometries of the corresponding supergravity backgrounds are different.
{}From the field theory point of view, now 
the Lagrangian contains a dimension 6 operator,
 and the resulting  theory is expected to
be non-renormalizable. In the spatial NCSYM theory it is important to have 
an infinite series of terms. If there are only a finite number of terms
it is hard to see why renormalization works.
Renormalizability in the light-like non-commutativity
seems to work in the same way as in the spatial NCSYM theory, in the sense 
that the divergences in the planar diagrams are taken care of as in the
SYM case. For the non-planar case, as long as 
$(p_0-p_1)^2 \neq 0$ there are no divergences, whereas
when $(p_0-p_1)^2 = 0$
we get the divergences
that are interpreted  as IR divergences \footnote{We would like
to thank J. Gomis for a discussion on this point.}.
It is worth noting that in a gauge $A_-=0$ the new interaction
(\ref{oop}) will always involve multiplicative factors $p_-$, which
may lead to an improvement of the IR behavior of non-planar diagrams.
Clearly, a more detailed analysis is needed in order 
to understand the structure of
perturbation theory for the ${\cal N}=4$ SYM theory with  light-like non-commutativity.


\subsection{M5 brane with light-like C field}

To find a solution representing an M5 brane in the presence
of a light-like C-field, we can proceed as above and boost the
 M5 brane solution with $C_{345}$ component \cite{MR}, in the direction
$x_3$. This is 
\be
ds^2_{11}=(kf)^{ {1\over 3}}\big[ {1\over f}\big( -d\tilde x_0^2
+dx_1^2+dx_2^2\big) +{1\over k}\big( d\tilde x_3^2+dx_4^2+dx_5^2\big)
+dr^2+r^2d\Omega_4^2\big]\ ,
\label{mff}
\ee
$$
f=1+{l_p^3R^3\over r^3}\ ,\ \ \ \ k=1+\cos^2\a {l_p^3 R^3\over r^3}\ ,
\ \ \ \ R^3={\pi N\over\cos\alpha }\ ,
$$
\be
dC_3=\sin\a\ df^{-1}\wedge d\tilde x_0\wedge dx_1\wedge dx_2+\cos\a\ 3R^3l_p^3\epsilon_4-
6\tan\a\  dk^{-1}\wedge d\tilde x_3\wedge dx_4\wedge dx_5\ ,
\label{cct}
\ee
where by $\epsilon_4$ we denote the volume form of the 4-sphere, and
we have made the Lorentz boost
\be
\tilde x_0=\cosh\g \ x_0-\sinh\g \ x_3\ ,\ \ \ \ \ 
\tilde x_3 =-\sinh\g \ x_0 +\cosh\g \ x_3\ .
\label{lll}
\ee
Now we take the limit $\g\to\infty $, $\a\to 0 $ 
with $e^\g \tan\a =b$=fixed. We get
$$
ds^2_{11}=f^{ -{1\over 3}}\big[  dx_+dx_-
+dx_1^2+dx_2^2+ dx_4^2 +dx_5^2
$$
\be
-\ { b^2\over  f} dx_-^2\big]
+f^{{2\over 3}}\big(dr^2+r^2d\Omega_4^2\big)\ ,
\label{nope}
\ee
\be
dC_3=b df^{-1}\wedge dx_-\wedge dx_1\wedge dx_2+
3R^3l_p^3\epsilon_4-
6b df^{-1}\wedge d x_-\wedge dx_4\wedge dx_5\ .
\ee
where we have redefined $x^+\to x^+ - b^2 x^-$.
This represents a gravitational wave moving parallel to the M5 brane.

The decoupling limit is taken by rescaling variables as follows:
\be
r=l_p^3 u^2\ ,\ \ \ b=l_p^{-3} \tilde b\ .
\label{decc}
\ee
and then taking $l_p\to 0 $ with fixed $u, R, \tilde b$.
We obtain
$$
ds^2_{11}=l_p^2 \bigg( {u^2\over (\pi N)^{{1\over 3}} }\big[  dx_+dx_- 
+ dx_1^2+dx_2^2+ dx_4^2 +dx_5^2- {\tilde b^2\over \pi N} u^6 dx_-^2\big]
$$
\be
+ (\pi N)^{1\over 3} \big[ {4du^2\over u^2}+ d\Omega_4^2 \big]\bigg)\  ,
\label{M5}
\ee
\be
dC_3=l_p^3\bigg({6\tilde bu^5\over \pi N} du\wedge dx_-\wedge dx_1\wedge dx_2+
3\pi N\epsilon_4-
{36\tilde b u^5\over \pi N} du\wedge d x_-\wedge dx_4\wedge dx_5\bigg)\ .
\label{C}
\ee

At low  energy (small $u$) the background (\ref{M5}), (\ref{C})
reduces to $AdS_7\times S^4$ as expected.
The curvature invariants are the same as in the $AdS_7\times S^4$ case
(again by virtue of the symmetry under rescalings of $\tilde b$, combined with a rescaling of coordinates).
The background  (\ref{M5}), (\ref{C}) provides a dual description 
of the $(0,2)$ theory perturbed by a dimension nine operator. 
This theory is conjectured to have a matrix-like description
as the quantum mechanics on the resolved moduli space of instantons.
The light-like $C$ field is interpreted as the resolution parameters
(the Fayet-Iliopoulos parameters
of the $0+1$ Yang-Mills theory) \cite{ABS} (see also \cite{Ganor}).

An alternative derivation of the solution (\ref{nope}) is by using as starting point the supergravity solution of M5-branes in the presence of C field with rank 4 \cite{AOS}
\bea
ds^2&=&h^{-{2\over 3}}
\left[f^{-{1\over 3}}\left(-dx_0^2+hdx_{1,2,3,4}^2+
h^2(dx_{5}-Cdx_0)^2\right)+
f^{2\over 3}(dr^2+r^2d\Omega^2_4)\right]\ ,
\cr
&&\cr
f&=&1+\frac{\pi N l_p^3}{\cos^2\theta r^3}\ ,\;\;\;\;\;\;\;h^{-1}=\sin^2\theta 
f^{-1}+\cos^2\theta \ ,\;\;\;\;C=\sin^2\theta f^{-1}, \cr
&&\cr
C_{012}&=&\cos\theta \sin\theta f^{-1}h\ ,\;\;\;\;\;\;\;\;\;\;
C_{345}=\tan\theta f^{-1}h\ , \cr
&&\cr
C_{034}&=&\cos\theta \sin\theta f^{-1}h\ ,\;\;\;\;\;\;\;\;\;\;
C_{125}=\tan\theta f^{-1}h \ .
\eea
Dimensional reduction along the
direction $x_5$ gives the D4 brane supergravity background in presence of
a B-field with magnetic components.
The infinite boost limit can be taken by introducing 
$$
\tilde x_0=x_0\cos\theta\ ,\ \ \ \ 
\tilde x_5={x_5 \over\cos\theta }\ ,\ \ \  \
$$
and taking the limit $\theta \to \pi/2$ with fixed
$\tilde x_0,\ \tilde x_5 ,\  r $ and {\it fixed} $l_P$.
In this way one reproduces the background (\ref{nope}).
The Dp brane backgrounds can then be obtained by
dimensional reduction along either of the coordinates
$(1,2,3,4)$ and T-dualities.

\subsection{NS5-branes with light-like RR fields}

By an S-duality transformation on the D5-brane solution, given by 
eq.~(\ref{bbk}) with $p=5$,  one 
finds the solution representing type IIB NS5-branes
in the presence of a light-like RR $A$ field:
\bea
ds^2&=&dx_{-}dx_{+}+\sum_{i=2}^{5}dx_i^2-{b^2\over f}dx^2_{-}+
f(dr^2+r^2d\Omega_3^2)\ ,
\cr
&&\cr
f&=&1+\frac{ N l_s^2}{r^2}\ ,\;\;\;\;\;\;\;e^{2\phi}=g_s^2f\ ,
\ \ \ l_s\equiv \sqrt{\a '}\ ,
\cr
&&\cr
A_{-2}&=&{b\over g_s}f^{-1}\ ,\;\;\;\;\;\;A_{-345}=-{b\over g_s}f^{-1}\ .
\label{wwq}
\eea
Using T-duality one can also find the type IIA NS5-branes with  light-like
RR  fields. T-duality in the direction $x_2$ gives 
an NS5-brane in the presence of light-like RR one- and 4-forms,
\be 
A_{-}= {b\over g_s}f^{-1}\ ,\ \ \ \ \   A_{-2345}=-{b\over g_s}f^{-1}\ ,
\label{ppo}
\ee
with the same metric and dilaton fields.
T-duality in the direction $x_3$ gives
an NS5-brane in the presence of light-like RR three-form, with components
\be 
A_{-23}= {b\over g_s}f^{-1}\ ,\ \ \ \ \   A_{-45}=-{b\over g_s}f^{-1}\ .
\label{ppoo}
\ee
The decoupling limit for these NS5-branes with light-like gauge fields is taken in the same
way as that for the usual NS5-branes \cite{agm},
 namely, $g_s\rightarrow 0$ and $l_s$=fixed, but in addition
we have to rescale ${\tilde b}=g_s b,\;r=g_sl_su$ with fixed $\tilde b$ and $u$.
Setting $u= N^{1\over 2}e^{z/r_0}$, with $r_0=l_s\sqrt{N}$,
we get for the type IIB NS5-branes (\ref{wwq})
\bea
ds^2&=&dx_{-}dx_{+}+\sum_{i=2}^{5}dx_i^2-{\tilde b}^2 e^{2z/r_0}dx_{-}^2+
dz^2+r_0^2d\Omega_3^2 \ ,
\cr
&&\cr
A_{-2}&=&{\tilde b}\ e^{2z/r_0 },\;\;\;\;A_{-345}=-{\tilde b}\ e^{2z/r_0 }\ ,
\;\;\;\;\phi=-{z\over r_0}\ .
\eea
 For the Type IIA  NS5-branes,  the metric
and dilaton are the same, and the gauge field components in
 eqs.~(\ref{ppo}), (\ref{ppoo}) become $A_-={\tilde b}\ e^{2z/r_0 }$, etc.
These backgrounds provide  dual descriptions of non-commutative little string
theories. The deformation parameters are the light-like RR fields.
 The phase structure of the theories is the same as 
the ordinary little string theories.
They are  
characterized by the linear dilaton behavior.

The Type IIB  NS5-brane
has a DLCQ description as a low energy SCFT of 
the Coulomb branch of a $1+1$ dimensional gauge theory 
\cite{set,ganset,ab}. In this case, the deformation parameters
are identified as mass parameters.
For the Type IIA NS5-branes,
the DLCQ deformation corresponds 
to turning on a
Fayet-Iliopoulos term \cite{ab} in the corresponding $1+1$ dimensional gauge
theory \cite{m1,m2}.

\section{NS5-branes in the presence of RR fields}

In this section we will study the theory on the Type II NS5-branes
in the presence of different RR field strengths, which can be either
electric or magnetic. The supergravity equations of 
motion require that
for NS5-branes in the presence of an
RR magnetic (electric) $(p+1)$-form there is also a RR electric (magnetic) $(5-p)$-form with $p=0,\cdots,5$. 
Therefore
the theory of NS5-branes in the presence of an electric (magnetic) $(p+1)$-form is
the same as the theory on the NS5-branes in the presence of a magnetic (electric)
RR $(5-p)$-form. 
For NS5-branes with a RR 3-form  there is no difference 
between electric and magnetic, so there is only one case to be studied.

The supergravity solution for NS5-branes in the presence of an RR $(p+1)$-
form is given by
\bea
ds^2&=&h^{-1/2}\left[-dx_0^2+\sum_{i=1}^{p}dx_{i}^2+h\sum_{j=p+1}^{5}dx_{j}^2
+f(dr^2+r^2d\Omega^2_3)\right]\ ,\cr
&&\cr
f&=&1+\frac{ N l_s^2}{\cos\theta\;r^2}\ ,\;\;\;\;\;\;\;\;h^{-1}=
\sin^2\theta f^{-1}+\cos^2\theta \ ,\cr
&&\cr
A_{0\cdots p}&=&{\sin\theta \over g_s} f^{-1}\ ,\;\;\;\;\;\;\;A_{(p+1)\cdots 
5}={\tan\theta \over g_s} f^{-1}h\ ,\;\;\;\;
e^{2\phi}=g_s^2f h^{(1-p)/2}\ .
\label{ffrr}
\eea
For $p=5$, $A_{(p+1)\cdots 5}$ denotes the RR scalar field.
A way to find these solutions is to start with  M5-branes in the presence of
a $C$ field (\ref{mff}), smeared in some transverse direction. By reducing on 
this transverse direction, 
one finds the type IIA NS5-branes with an electric RR 3-form.
Other solutions are generated by T-duality.

The decoupling limit of the above supergravity solution can be defined as
the limit  $l_s\rightarrow 0$, keeping the following quantities fixed:
\def\af{ \alpha'_{\rm eff}}
\def\lef{ l_{\rm eff} }
\be
\af ={l_s^2\over \cos\theta}\ ,\;\;\;\;\;\;u={r\over l_s^2}\ ,\;\;\;\;\;\;
{\tilde g}\ \lef^{p-3}=g_sl_s^{p-3}\ ,
\label{yyyy}
\ee
\be
{\tilde x}_{0,\cdots,p}={1\over \lef }x_{0,\cdots,p}\ \ ,\;\;\;\;\;
{\tilde x}_{(p+1),\cdots,5}={\lef \over l_s^2}x_{(p+1),\cdots,5}\ ,
\ \ \ \ \lef\equiv\sqrt{\af }\ .
\label{ddee}
\ee
In this limit the supergravity solution becomes
\bea
l_s^{-2}ds^2&=&(1+a^2u^2)^{1/2}\left[-d{\tilde x}_0^2+\sum_{i=1}^{p}
d{\tilde x}_{i}^2+\frac{\sum_{j=p+1}^{5}d{\tilde x}_{j}^2}{1+a^2u^2}
+{ N\over u^2}(du^2+u^2d\Omega^2_3)\right]\ ,\cr
&&\cr
A_{0\cdots p}&=&{l_s^{(p+1)} \over {\tilde g}} a^2u^2\ \ ,
\;\;\;\;\;\;\;\;\;\;
\;\;\;\;\;\;
A_{(p+1)\cdots 5}={l_s^{(5-p)}\over {\tilde g}}\;\frac{a^2u^2}{1+a^2u^2}\ ,
\cr
&&\cr
e^{2\phi}&=&{\tilde g}^2 \frac{(1+a^2u^2)^{(p-1)/2}}{a^2u^2}\ ,\;\;\;\;
a^2={\af   \over N}\ .
\label{rrrr}
\eea
These backgrounds provide a supergravity dual description for the 
ODp theories investigated in \cite{GMSS}\footnote{The supergravity 
description of ODp with  $p=1,2$ and $p=2,3$ has  also been 
considered in \cite{har} and \cite{A}, respectively.}
(with coupling $g_{YM}^2={\tilde g}\ \lef^{p-3}$).
The scalar curvature of the metric is given by
\be
l_s^2{\cal R}= {1\over N}\;\frac{c_1+c_2 a^2 u^2 +c_3 a^4 u^4}
{(1+a^2u^2)^{5\over 2} }\ ,
\ee
where $c_1,c_2,c_3$ are numerical constants depending only on $p$.
Therefore for large $N$ the curvature is small and  one can trust 
the supergravity description. 
 
As an application, let us now consider the absorption cross 
section of polarized gravitons. This 
 calculation has already been done for the type
IIA NS5-branes in the presence of an RR 3-form and type IIB NS5-branes
in the presence of a magnetic RR 2-form in \cite{{A},{AIO}}, which 
correspond to the supergravity dual of OD2 and OD3 theories, respectively.
In general one can show that in these backgrounds
the scattering potential for a graviton polarized along the brane directions
is 
\be
V(\rho)=-1+({3\over 4}-\omega^2 R^2)\;{1\over \rho^2}\ , \;\;\;\;
R^2={N l_s^2\over \cos\theta}\ ,
\ee
where $\rho=\omega r$ and $\omega$ is  the energy of incoming waves.
Therefore we see that after the decoupling limit  
the absorption cross section can be nonzero only 
for  waves with  energy $\omega^2 $ larger than $\sim {1\over N \af }$. 
Essentially the same effect appears in the little string theories. 
Following \cite{MS}, one can see that these theories have a mass gap 
of order $M^2_{\rm gap}\sim {1\over N \af }$~.
To compare the decoupling limit (\ref{hhhh}) 
with  that of ordinary little string theories, 
it is convenient to 
describe the decoupling limit in terms of $g_s$. 
For $p\leq 2$,
one can take the decoupling limit of the NS5-branes
in the presence of electric $(p+1)$-form  as follows
\be
g_s\rightarrow 0\ ,\;\;\;\;{\tilde g}^{2\over 3-p}=
{g_s^{2\over 3-p }\over \cos\theta}\ ,\;\;\;\;r=g_s^{1\over 3-p}l_s \  u\ .
\ee  
with fixed $l_s , u, {\tilde g}$. In this limit the supergravity 
background (\ref{ffrr}) reduces 
to the same expression (\ref{rrrr}). This description is
equivalent to a rescaling of the coordinates. 
In this way one has $g_s\rightarrow 0$ and
$l_s$ fixed, as in the little string theory.

The ODp theories have all the same physics at low energies: 
for odd $p$ they
flow to  SYM theory in (5+1)-dimension in the IR; for even $p$, the
theories flow to a fixed point in the  IR described by the (0,2) conformal 
theory. 

In the ultraviolet  regime, where the effects of 
nonzero RR fields become important, the different ODp theories exhibit  
different behaviors, according to the value of $p$. 

For $p\leq 2$ case, the string coupling $e^\phi $ in eq.~(\ref{rrrr}) 
is small in the ultraviolet 
regime and one can trust the supergravity solution. In this region  
$u\gg a^{-1}$
the  NS5 brane supergravity reduces to a metric
describing ordinary Dp-branes  smeared in $5-p$ directions. 
In the particular case of the OD0 theory, the supergravity solution 
(\ref{rrrr})  provides a supergravity description of a DLCQ compactification
of M-theory with $N$ units of DLCQ momentum, in the presence of a transverse 
M5-brane. The relation between M-theory and type IIA parameters is as follows:
\be
\af {\tilde g}^{2/3}=M_{\rm eff}^{-2} \ ,\;\;\;\;\;\;\af {\tilde g}^2=R_{11}^2\ ,
\label{LIFT}
\ee
where $M_{\rm eff}$ if the effective eleven-dimensional Planck mass.

For $p=3$ the dilaton in (\ref{rrrr}) 
is constant at large $u$, i.e.  $e^{\phi}={\tilde g}$. 
For ${\tilde g}\ll 1$ the theory can be described by smeared D3-branes, 
while for ${\tilde g}\gg 1$ we have to use the S-dual 
picture describing D5-branes in the presence of a magnetic $B$ field with 
rank two.
Therefore, in the UV regime, strongly coupled OD3 theory and large $N$ 
5+1 dimensional NCSYM  theory 
exhibit a similar behavior. Note that under S-duality the parameters
of the theory change as
\be
\af \to {\tilde g} \af \ ,\;\;\;\;\;{\tilde g}\to {1\over {\tilde g}}\ ,
\label{hhhh}
\ee 

For $p= 4$ the dilaton is large at $u\gg a^{-1}$.
In this case it means that the proper supergravity description is in terms
of eleven-dimensional supergravity. 
From M-theory point of view, the supergravity solution is
the bound state of two M5-branes in the directions
(0,1,2,3,4,5) and (0,1,2,3,4,6) in the decoupling limit. 

For $p=5$, the dilaton is also large  at $u\gg a^{-1}$.
The S-dual picture is not useful, since the transformed
dilaton field $\phi '$ is also large in this regime. 
Indeed, due to the  nonzero
RR 0-form (of order one), under S-duality we find
$e^{\phi'} \sim \tilde g a u$.
This is in agreement with the discussion of \cite{GMSS}. It
 can be understood from the fact that, under S-duality 
the system maps to a similar configuration of NS5-branes in the presence of
electric RR 6-form.
If $N$ is the  number of NS5-branes and $M$ the charge induced by
RR 6-form one has the relation \cite{H}
${1\over \cos \theta}=g_s{M \over N}$.
In the decoupling limit (\ref{yyyy}), one obtains
\be
{\tilde g}={N\over M}\ .
\ee
A similar relation is found for D1-branes in the presence of electric
B field \cite{{GMSS},{km}}. 


{} A T-duality transformation on the background (\ref{rrrr}) 
implies the following relation between the parameters of ODp theory and OD(p-1) theory:
\be
R\rightarrow {\af \over R}\ ,\;\;\;\;\;\;{\tilde g}^2\rightarrow {\af\over R^2}
{\tilde g}^2\ .
\label{uuuu}
\ee 
Thus, from eqs.~(\ref{LIFT}), (\ref{hhhh}) and (\ref{uuuu}), we see  that
the parameters of OM, NCOS and  ODp theories are 
related in the same way as
the corresponding  parameters of type IIA, type IIB and M-theory, as expected.

\vskip 1cm  

{\bf Acknowledgement}:
We wish to thank O. Aharony, J. Gomis and  A.~Tseytlin for valuable discussions.
Y. O.  would like to thank the USC/Caltech center for theoretical
physics for hospitality during
the course of this work. J. R. would like to thank CERN for hospitality.

\end{document}